# Flexible strained membranes of multiferroic TbMnO$_3$

H. Shi,[1] F. Ringe,[1] D. Wang,[2] O. Moran,[3] K. Nayak,[1] A. K. Jaiswal,[1] M. Le Tacon,[1] and D. Fuchs[1,*]

[1]*Karlsruhe Institute of Technology, Institute for Quantum Materials and Technologies,*

*Kaiserstr. 12, 76131 Karlsruhe, Germany;*

[2]*Karlsruhe Institute of Technology, Institute of Nanotechnology and Karlsruhe Nano Micro Facility,*

*Kaiserstr. 12, 76131 Karlsruhe, Germany;*

[3]*Universidad Nacional de Colombia, Sede Medellín-Facultad de Ciencias, Departamento de Física, Advanced Oxides Group, Carrera 65 No. 59A-110, Medellín, 050034, Colombia;*

The multiferroic properties of TbMnO$_3$ demonstrate high versatility under applied pressure, making the material potentially suitable for use in flexible electronics. Here, we report on the preparation of elastic freestanding TbMnO$_3$ membranes with dominant (001) or (010) crystallographic out-of-plane orientation. Membranes with thickness of 20 nm display orthorhombic bulk-like relaxed lattice parameters with strong suppression of twinning for the (010) oriented membranes. Strain in flexible membranes was tuned by using a commercial strain cell device and characterized by Raman spectroscopy. The $B_{1g}$ out-of-phase oxygen-stretching mode, representative for the Mn-O bond distance, systematically shifts to lower energy with increasing strain ($\varepsilon_{max} \approx 0.5\%$). The flexibility and elastic properties of the membranes allow for specific manipulation of the multiferroic state by strain, whereas the choice of the crystallographic orientation gives possibility for an in- or out-of-plane electric polarization.

Magnetoelectric multiferroic (MF) materials have garnered significant interest in the field of next-generation spintronics, owing to their potential to revolutionize the way electronic devices consume energy [1–4]. The unique feature of these materials is their ability to couple magnetism and ferroelectricity, enabling the control of magnetic states through electric fields and vice versa. This coupling offers a substantial opportunity for reducing energy consumption in electronic devices. Unlike type-I MFs, which exhibit relatively weak coupling between their magnetic and ferroelectric states - despite often having ordering temperatures above room temperature - type-II MFs demonstrate much stronger magnetoelectric coupling. This enhanced coupling is due to the intrinsic mutual entanglement between the magnetic and ferroelectric states, making type-II MFs particularly promising for the electric field control of magnetization.

In most of the type-II MFs, ferroelectricity appears in conjunction with a spiraling or cycloid magnetic phase. One of the most prominent examples for this is the orthorhombic (*Pbnm*) perovskite TbMnO$_3$ (TMO).[5] Spin frustration leads to a complex magnetic order that breaks inversion symmetry.[6] Below $T_{N1} = 41$ K, the magnetic structure of TMO adopts a sinusoidal spin-density wave configuration, wherein all spins are aligned parallel to the *b*-axis. In this arrangement, the magnitude of the local magnetic moments varies periodically along the wave propagation direction. This centrosymmetric structure becomes ferroelectric below $T_{N2} = 28$ K, where a non-centrosymmetric cycloid spiral spin wave in the *bc*-plane develops. The ferroelectric polarization *P* is directed along the *c*-axis. The microscopic mechanism for *P* in TMO is related to the spin-orbit and Dzyaloshinski-Moriya interaction.[7]

Under pressure, TMO exhibits a flop of *P* towards the *a*-axis direction and a giant spin-driven ferroelectric polarization $P \approx 1$ μC/cm$^2$, which is about one order of magnitude larger compared to that observed at ambient pressure.[8,9] Applying pressure alters interatomic distances, thereby affecting spin frustration by modifying the subtle equilibrium of magnetic interactions[10]. Very similar to that, in thin rare-earth manganite films, epitaxial strain alters magnetic state as well.[11–13] Single-phase epitaxial films have been grown successfully on various perovskite-like single-crystalline substrates.[13,14] Depending on the strain state, TMO films can adopt either a spin-spiral-induced ferroelectric ground state as in the bulk, or an E-type antiferromagnetic ground state with large *P*.[13]

Therefore, strain seems to be very efficient in manipulating the magnetic state and hence *P* in TMO. However, strain state is fixed when films are grown on a specific substrate and cannot easily be varied to tune *P*. In addition, to further increase *P* by increasing epitaxial strain, *e. g.*, by choosing substrate materials with larger lattice mismatch, usually partial strain relaxation and the formation of lattice defects occurs which lowers the crystalline quality and MF properties of the film.[15] To address these deficiencies and explore the strain-dependence and tunability of MF properties, along with assessing their suitability for flexible electronics, employing freestanding (FS) TMO membranes offers a perfect ground.[16] As recently reported, FS perovskite-like oxide thin films can be produced with a high degree of crystallinity and potential to withstand large tensile strain.[17–21] Most of these FS films have been prepared by chemical approach using a sacrificial layer which can be dissolved in an appropriate etchant as for example the water-soluble oxide (Sr$_{1-x}$Ca$_x$)$_3$Al$_2$O$_6$.[22,23] The cubic crystal structure (*Pa*-3) has lattice parameter of $a/4 = 3.961$ Å to – 3.814 Å for $x = 0 – 1$, respectively, and therefore shows good crystalline compatibility and lattice matching to many functional perovskites.

In this letter, we report on the successful preparation of differently oriented TMO membranes by lattice mismatch engineering and the use of water-soluble epitaxial Sr$_3$Al$_3$O$_6$ (SAO) and Ca$_3$Al$_2$O$_6$ (CAO) sacrificial layers and demonstrate flexible state of strain of the membranes. Specifically, TMO FS films were grown with dominant *b*- or *c*-axis orientation, where *P* is aligned in- or perpendicular to the film plane, respectively. The crystal structure and lattice parameters were found to be the same as those of the bulk material, indicating a strain-free state of the FS films. Flexible state of uniaxial tensile strain is demonstrated by using commercial strain-cell.

Epitaxial TMO, SAO, and CAO thin films with a film thickness $t = 20$ nm were grown by pulsed laser deposition (PLD) technique. For the laser ablation, stoichiometric targets were prepared by standard solid-state synthesis. The deposition parameters (see *Supplementary Material*) were optimized with respect to stoichiometry and crystallinity. The bulk lattice parameters of TMO ($a = 5.303$ Å, $b = 5.8307$ Å, and $c = 7.4146$ Å) result in a different surface unit cell for (010) and (001) orientation, see Fig. 1a,b. Therefore, by variation of the lattice mismatch between the surface cell of TMO and the substrate material, a preferential growth orientation can be obtained. In Fig. 1c we have demonstrated the epitaxial growth of TMO on various perovskite-like substrate materials. The pseudo-cubic lattice parameters $a_{pc}$ and surface cell $a_{pc}^2$ for the used substrates are: 3.70 Å, 13.70 Å$^2$ for (010-YAlO$_3$), 3.78 Å, 14.29 Å$^2$ for (001-LaAlO$_3$), 3.87 Å, 14.97 Å$^2$ for (001-(LaAlO$_3$)$_{0.3}$(Sr$_2$AlTaO$_6$)$_{0.7}$), *i. e.*, (001-LSAT), and 3.91 Å, 15.29 Å$^2$ for (001-SrTiO$_3$). Here, $a_{pc}$ was deduced from the orthorhombic/rhombohedral lattice parameters by $a_{pc} = (a_o/\sqrt{2}+b_o/\sqrt{2}+c_o/2)/3$. The surface cell of (010) and (001) TMO amounts to 13.90 Å$^2$ and 15.46 Å$^2$, respectively. Therefore, best lattice and symmetry matching for the epitaxial growth of (010) and (001) oriented films is obtained on YAlO$_3$ (010) and SrTiO$_3$ (001), respectively. The 2θ x-ray-diffraction spectra shown in Fig. 1c from top to bottom indicate the evolution of epitaxy for TMO with increasing substrate surface cell. The growth orientation of TMO systematically changes from (010) orientation on (010) YAlO$_3$ (YAO), which is mainly triggered by the substrate surface symmetry, to a (001) orientation on (001) SrTiO$_3$ (STO) where substrate surface symmetry and cell size matches best. In-between, a mixed (110)/(001) growth orientation on (001) LaAlO$_3$ and LSAT occurs.

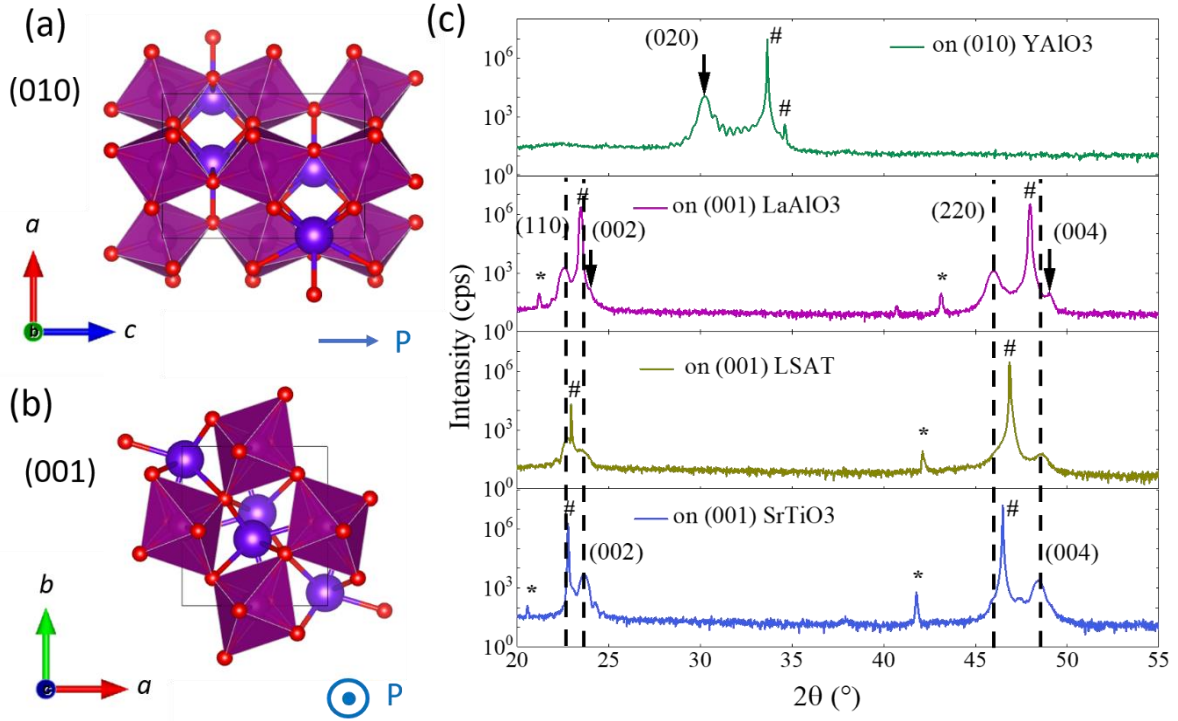

FIG. 1. Crystal structure of TMO with projection along the (010) (a) and the (001) direction (b). MnO$_6$ octahedra are indicated. Tb and O atoms are shown by purple and red spheres, respectively. The surface unit cell and the direction of the ferroelectric polarization $P$ are indicated alike. (c) 2θ x-ray diffraction (Cu$K_\alpha$ radiation) for TMO films ($t = 20$ nm) on various substrate materials. Substrate surface cell increases from top to bottom. Substrate peaks due to Cu$K_\alpha$(#) and Cu$K_\beta$(*) are marked.

Consequently, for the preferential growth of (010) and (001) TMO films (010) YAO and (001) STO substrates were used, respectively. The preparation of the CAO/SAO sacrificial layer was carried out in-situ before the TMO deposition. We limited $t$ to 20 nm in order to avoid distinct lattice relaxation and defect formation of CAO or SAO. The good water solubility of CAO/SAO is related to the good hydration of the oxygen in the $AlO_4$ groups which form a closed ring structure within the cubic (Pa-3) unit cell,[24,25] see Fig. 2a. For $t \approx 20$ nm the etching time in distilled water amounts to about 1 h and rapidly increases for smaller $t$. With respect to the lattice matching, CAO ($a/4 = 3.81$ Å) was used for the growth on YAO ($a_{pc} \approx 3.70$ Å) and SAO ($a/4 = 3.97$ Å) for the deposition on STO ($a = 3.905$ Å). According to the "block by block" growth of perovskite related structures, the cubic cell of CAO grows with (110) orientation on (010) YAO, whereas SAO grows with (001) orientation on (001) STO. The epitaxial growth of the TMO heterostructures is documented by 2θ x-ray diffraction scans in Fig. 2b.

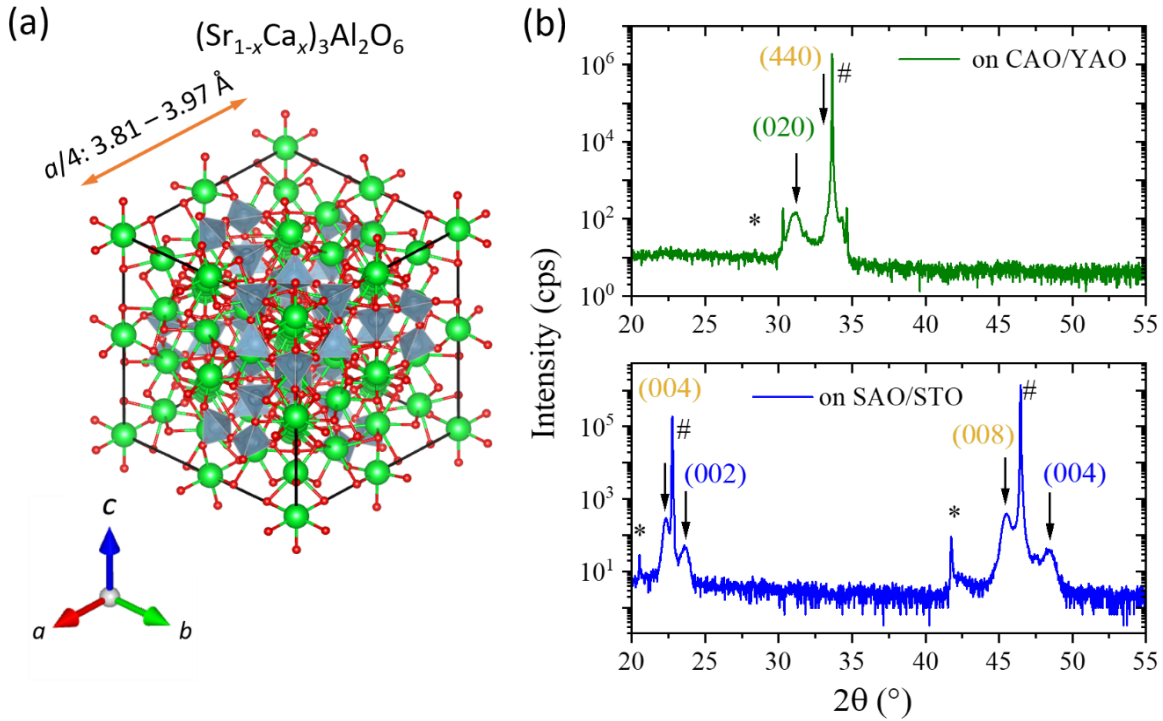

FIG. 2. (a) Cubic (Pa-3) crystal structure of $(Sr_{1-x}Ca_x)_3Al_2O_6$. The $AlO_4$ tetrahedra are shown in grey color. When viewing along the (111) direction, rings comprising 6 $AlO_4$ groups which are responsible for the good water solubility can be realized. Ca/Sr is indicated by green spheres. Lattice parameter increases from $a/4 = 3.81$ Å for $x = 1$ (CAO) to 3.97 Å for $x = 0$ (SAO). (b) 2θ x-ray diffraction for TMO films on CAO/YAO (010) and on SAO/STO (001). Substrate peaks due to $CuK_\alpha$(#) and $CuK_\beta$(*) are marked. CAO/SAO and TMO film reflections are indicated in orange and green or blue, respectively.

The transfer process of complex oxide films represents the largest difficulty for obtaining large-area and crack-free films and different routes have been studied intensively.[18] To avoid heavy wrinkling of the FS films, lattice-mismatch between film and sacrificial layer and internal stress in the polymer support should be reduced as much as possible. Here, we used the polymer (poly)dimethyl siloxane (PDMS) as support material. The PDMS support was attached to the TMO film surface and watered in distilled $H_2O$ for about 2h after which the CAO/SAO sacrificial layer was usually completely dissolved. So, the obtained films on PDMS could be easily transferred to other support materials by using an adhesive tape (PRO Tapes – Nitto SPV224). In Fig.3a we have displayed a 5×5 mm² TMO film after transfer process.

High resolution (HR) scanning transmission electron microscopy (STEM) documents the orthorhombic structure, stoichiometric composition, and high crystalline quality of the TMO membranes (Fig. 3b,c). 2θ-x-ray diffraction scans display dominant (010) and (001) growth orientation of the FS TMO films grown on CAO/YAO (010) and SAO/STO (001), respectively (see Fig. 3d,e). FS films also display minor contribution of (100) and (110) phase which could not be seen before the etching process due to the presence of the YAO and STO substrate peaks, respectively (*cf.* Fig. 2b). The in-plane texture was probed by φ-scans which verify two-fold and four-fold symmetry of orthorhombic (021) and (010) reflections, respectively. This indicates prominent untwinned growth on CAO/YAO (010) and a twinned growth on SAO/STO (001) in the film plane (see also *Supplementary Material*). Usually, epitaxial growth of orthorhombic TMO on a surface showing four-fold symmetry such as STO, SAO or CAO results in a twinning of TMO. Depending on the lattice mismatch this may be restricted to an in-plane or out-of-plane twinning but can also result in the presence of all three-principle film axis along one substrate direction. The observed dominant TMO grain orientations are visualized in Fig. 3 alike. The FS films display nearly completely strain relaxed orthorhombic state with bulk like lattice parameters: $a = 5.82$ Å, $b = 5.43$ Å, and $c = 7.49$ Å.

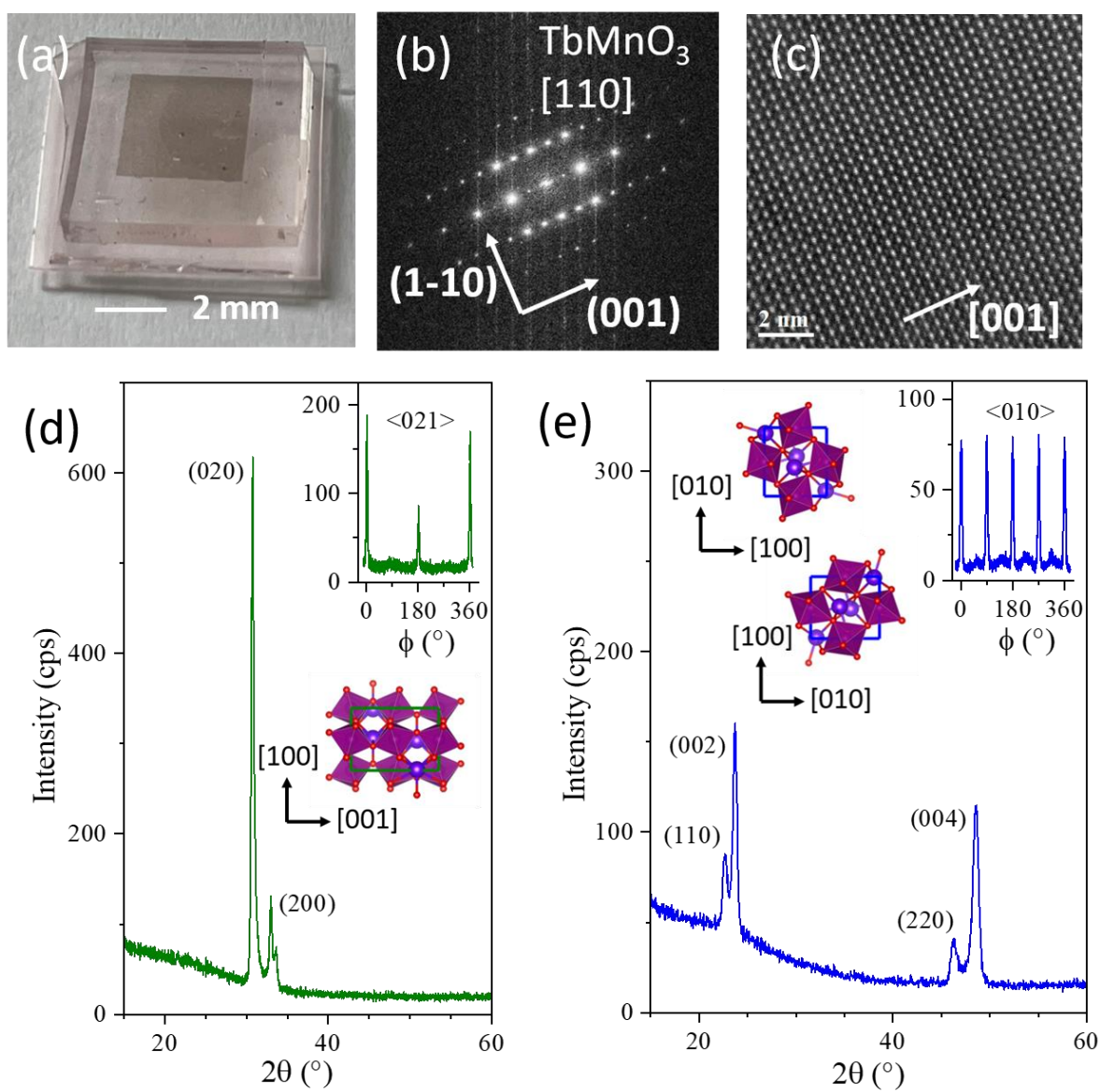

FIG.3. (a) TMO film (5×5 mm$^2$) after transfer to PDMS support. (b) Plan view TEM electron diffraction and (c) HR-STEM along [110] zone axis of (110) minor phase of TMO FS film obtained from SAO/STO. (d) 2θ-scan of FS TMO film grown on CAO/YAO. Film shows dominant (010) orientation. The ϕ-scan on the (021) reflection of (010) oriented grains (inset) documents two-fold symmetry and hence untwinned growth of TMO. (e) 2θ-scan of FS TMO film obtained from SAO/STO. FS film displays dominant (001)-orientation. ϕ-scan on the <010> reflection of (110) oriented grains (inset) shows four-fold symmetry indicating in-plane twinning of the TMO film. Top view of the observed dominant TMO grain orientation is indicated for both films.

To probe the flexibility of the freestanding TMO membranes, continuous uniaxial strain was applied to the membranes by using a commercial strain-cell (Razorbill CS200T). To this purpose, exfoliation of the TMO films in water was done without using PDMS support. The dispersed film material was directly picked out from the water with a micro-tweezer and transferred to a split Si-wafer which was glued to the left and right sample plate of the strain-cell, see Fig. 4a. The membranes were fixed to the Si only by adhesion. The dispersion in water also allowed us to obtain furled tube-like film flakes with similar diameter of about 3.5 µm, see also Fig. 4a. The self-rolling-up process of the TMO membranes is likely triggered by the release of residual internal-strain gradient or surface tension.[26,27] These tube-like flakes are highly suitable to probe film flexibility alike. The strong bending of the TMO tubes results in a tensile strain on the convex surface which can be estimated by the film thickness b and bending radius R ($\varepsilon \approx 0.57$ %),[28] see Fig. 4b. In contrast, the uniaxial strain generated by the strain cell is given by the relative change of the sample length, ΔL/L. Ideally, ΔL and L correspond to the increase of the gap, Δ, and the initial gap-distance $d_0 \approx 5.45$ µm, respectively. The variation of strain in TMO was analyzed by the frequency change of the Raman active $B_{1g}$ in-plane oxygen-stretching mode, see Fig. 4c, which is highly sensitive to changes of the Mn-O bond distance.[29,30] The Raman spectra were recorded at $T = 120$ K, for (001)-oriented TMO membranes on Si (relaxed TMO), tubes (bended TMO) and flakes across the gap (tensile strained TMO), see Fig. 4c, and are well consistent with spectra reported in literature.[31–33] Beside the $B_{1g}$ mode at 618 cm$^{-1}$ we also indicated the $A_g$ out-of-phase MnO$_6$ rotation mode at 382 cm$^{-1}$ (see *Supplementary Material*). By increasing the gap distance and hence tensile strain, the $B_{1g}$ mode shifts to lower wavenumber as expected from measurements under pressure, where an upward shift of the mode is observed.[31,33] Symmetry of the $B_{1g}$ mode with respect to the *a*- and *b*-axis direction as well as the in-plane *a-b*-twinning of the membranes makes measurements rather insensitive to strain-direction. Increasing Δ to 0.45 µm results in a cracking of the membrane and the subsequent increase of the $B_{1g}$ wavenumber towards its initial value. For a rough estimation of the strain state the right scale of Fig. 4d shows the mean tensile strain ($\Delta a/a + \Delta b/b$)/2 deduced from hydrostatic pressure experiments on TMO single crystals, which is expected for the observed $B_{1g}$-shift.[33] The estimation gives reasonable values for the TMO tube and indicates only minor strain transfer from the cell to the film due to insufficient adhesion of the flakes on Si. Although the maximum strain achieved experimentally is only moderate, the evident flexibility and elastic properties of the membranes are promising. To further assess the strength of the magnetoelectric coupling in strained TMO membranes requires further investigation, through e.g. the study of Raman active electromagnon modes below $T_{N2}$[34].

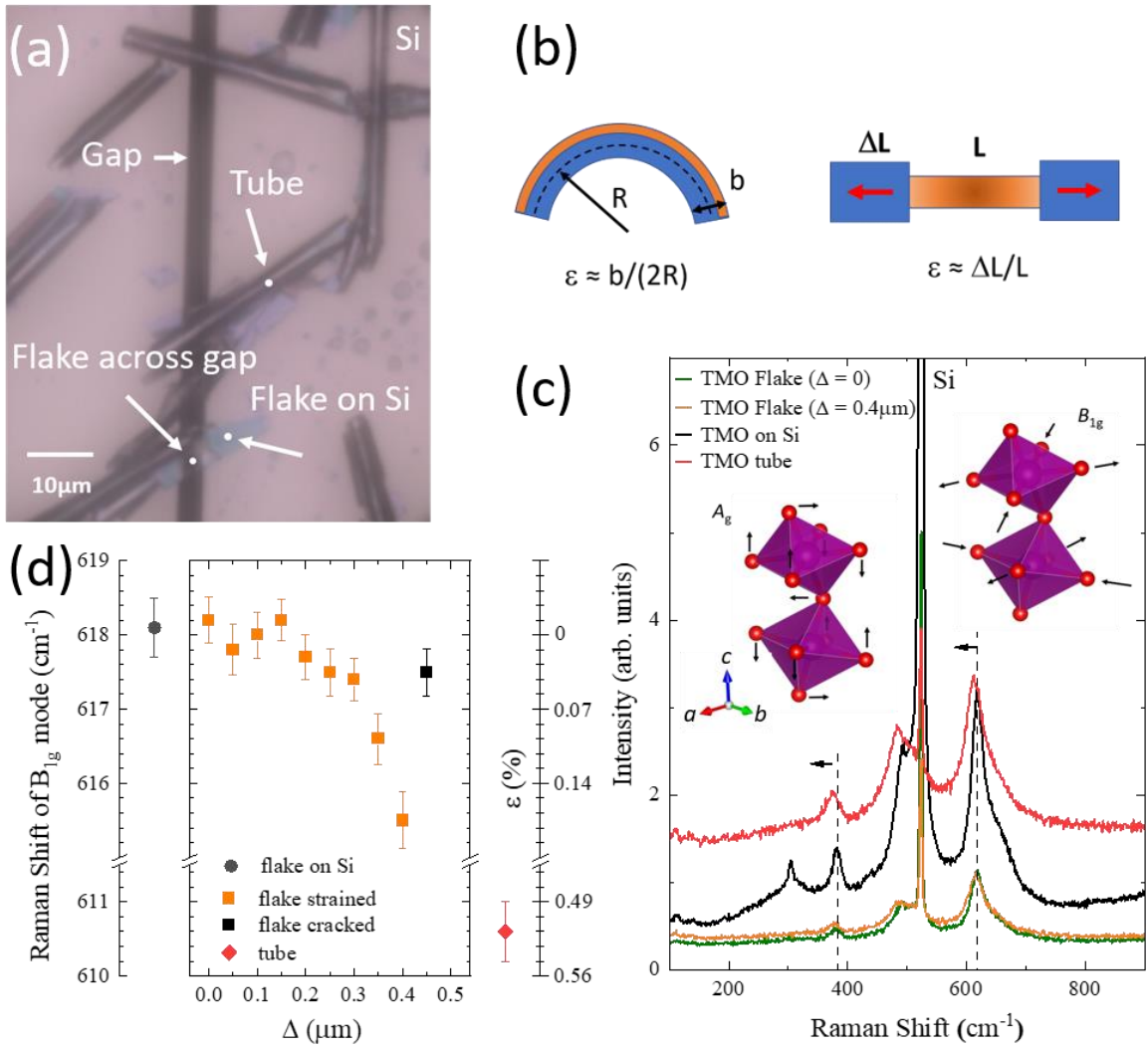

FIG.4. (a) TMO flakes and tubes with (010) surface orientation on a split Si-wafer which has been mounted to a commercial strain cell. Spot-positions were Raman measurements were taken are indicated. (b) Strain generated by bending (left) or uniaxial elongation (right). (c) Raman spectra of the TMO FS film material shown in (a). The $B_{1g}$ stretching and the $A_g$ rotation modes are indicated. The dashed lines show initial mode position which shifts to lower wavenumber with increasing tensile strain. (d) Raman shift of the $B_{1g}$ mode for the TMO flake on Si, the strained and cracked flake, and the tube. Right scale indicates estimated tensile strain, see text.

In summary, we have demonstrated the preparation of dominantly (001) and (010) oriented TMO membranes. The membranes display flexible and elastic properties which will allow specific manipulation of the multiferroic state by strain. In addition, the disposability of (001) and (010) oriented membranes provides avenue for out-of-plane and in-plane electric polarization, respectively.

See the *Supplementary Material* for details of the film preparation, characterization of structural and magnetic properties and Raman measurements. Please refer to supporting data guide-lines at https://aip.scitation.org/apl/authors/manuscript.

**Acknowledgement**

We are grateful to R. Thelen and the Karlsruhe Nano-Micro Facility (KNMF) for technical support. We thank J. Schubert, D. M. Buca, and K. Palmen from the Peter Grünberg Institute, Forschungszentrum Jülich, for carrying out Rutherford backscattering spectrometry. A. K. J. acknowledges financial support from the European Union's Framework Programme for Research and Innovation, Horizon 2020, under the Marie Skłodowska-Curie grant agreement No. 847471 (QUSTEC). H. Shi thanks F. Henßler for kindly help on the strain measurement and also acknowledges M. Graf von Westarp and A. von Ungern-Sternberg Schwark for useful discussions on data analysis. O.M. acknowledges financial support from the Universidad Nacional de Colombia. K. N. thanks German Academic Exchange Service (DAAD) for financial support.

**Author Declarations**

The authors have no conflicts to disclose.

**Data Availability**

The data that support the findings of this study are available from the corresponding author upon reasonable request.

**References**

[1] H. Béa, M. Gajek, M. Bibes, and A. Barthélémy, "Spintronics with multiferroics," Journal of Physics: Condensed Matter **20**(43), 434221 (2008).

[2] N.A. Spaldin, S.-W. Cheong, and R. Ramesh, "Multiferroics: Past, present, and future," Phys Today **63**(10), 38–43 (2010).

[3] S. Manipatruni, D.E. Nikonov, C.-C. Lin, T.A. Gosavi, H. Liu, B. Prasad, Y.-L. Huang, E. Bonturim, R. Ramesh, and I.A. Young, "Scalable energy-efficient magnetoelectric spin–orbit logic," Nature **565**(7737), 35–42 (2019).

[4] J.K. Jose, and R. Balakrishnan, "Multiferroics for Spintronic Applications," in *Nanotechnology in Electronics*, (2023), pp. 301–316.

[5] T. Kimura, T. Goto, H. Shintani, K. Ishizaka, T. Arima, and Y. Tokura, "Magnetic control of ferroelectric polarization," Nature **426**(6962), 55–58 (2003).

[6] M. Mostovoy, "Ferroelectricity in Spiral Magnets," Phys Rev Lett **96**(6), 67601 (2006).

[7] H. Katsura, N. Nagaosa, and A. V Balatsky, "Spin Current and Magnetoelectric Effect in Noncollinear Magnets," Phys Rev Lett **95**(5), 57205 (2005).

[8] T. Aoyama, K. Yamauchi, A. Iyama, S. Picozzi, K. Shimizu, and T. Kimura, "Giant spin-driven ferroelectric polarization in TbMnO3 under high pressure," Nat Commun **5**(1), 4927 (2014).

[9] T. Aoyama, A. Iyama, K. Shimizu, and T. Kimura, "Magnetic-field-induced ferroelectric polarization flop under pressure in TbMnO3," J Phys Conf Ser **592**(1), 012118 (2015).

[10] O.L. Makarova, I. Mirebeau, S.E. Kichanov, J. Rodriguez-Carvajal, and A. Forget, "Pressure-induced change in the magnetic ordering of TbMnO$_{3}$," Phys Rev B **84**(2), 20408 (2011).

[11] H. Wadati, J. Okamoto, M. Garganourakis, V. Scagnoli, U. Staub, Y. Yamasaki, H. Nakao, Y. Murakami, M. Mochizuki, M. Nakamura, M. Kawasaki, and Y. Tokura, "Origin of the Large Polarization

in Multiferroic ${\mathrm{YMnO}}_{3}$ Thin Films Revealed by Soft- and Hard-X-Ray Diffraction," Phys Rev Lett **108**(4), 47203 (2012).

[12] F. Jiménez-Villacorta, J.A. Gallastegui, I. Fina, X. Marti, and J. Fontcuberta, "Strain-driven transition from $E$-type to $A$-type magnetic order in YMnO${}_{3}$ epitaxial films," Phys Rev B **86**(2), 24420 (2012).

[13] K. Shimamoto, S. Mukherjee, S. Manz, J.S. White, M. Trassin, M. Kenzelmann, L. Chapon, T. Lippert, M. Fiebig, C.W. Schneider, and C. Niedermayer, "Tuning the multiferroic mechanisms of TbMnO3 by epitaxial strain," Sci Rep **7**(1), 44753 (2017).

[14] Y. Cui, C. Wang, and B. Cao, "TbMnO3 epitaxial thin films by pulsed-laser deposition," Solid State Commun **133**(10), 641–645 (2005).

[15] M.A. Carpenter, D. Pesquera, D. O'Flynn, G. Balakrishnan, N. Mufti, A.A. Nugroho, T.T.M. Palstra, M. Mihalik, M. Mihalik, M. Zentková, A. Almeida, J.A. Moreira, R. Vilarinho, and D. Meier, "Strain relaxation dynamics of multiferroic orthorhombic manganites," Journal of Physics: Condensed Matter **33**(12), 125402 (2021).

[16] S. Gupta, W.T. Navaraj, L. Lorenzelli, and R. Dahiya, "Ultra-thin chips for high-performance flexible electronics," Npj Flexible Electronics **2**(1), 8 (2018).

[17] S.S. Hong, M. Gu, M. Verma, V. Harbola, B.Y. Wang, D. Lu, A. Vailionis, Y. Hikita, R. Pentcheva, J.M. Rondinelli, and H.Y. Hwang, "Extreme tensile strain states in La0.7Ca0.3MnO3 membranes," Science (1979) **368**(6486), 71–76 (2020).

[18] F.M. Chiabrera, S. Yun, Y. Li, R.T. Dahm, H. Zhang, C.K.R. Kirchert, D. V Christensen, F. Trier, T.S. Jespersen, and N. Pryds, "Freestanding Perovskite Oxide Films: Synthesis, Challenges, and Properties," Ann Phys **534**(9), 2200084 (2022).

[19] L. Han, G. Dong, M. Liu, and Y. Nie, "Freestanding Perovskite Oxide Membranes: A New Playground for Novel Ferroic Properties and Applications," Adv Funct Mater **n/a**(n/a), 2309543 (2023).

[20] J. Ji, S. Park, H. Do, and H.S. Kum, "A review on recent advances in fabricating freestanding single-crystalline complex-oxide membranes and its applications," Phys Scr **98**(5), 052002 (2023).

[21] N. Pryds, D.-S. Park, T.S. Jespersen, and S. Yun, "Twisted oxide membranes: A perspective," APL Mater **12**(1), 010901 (2024).

[22] D. Lu, D.J. Baek, S. Sae Hong, L.F. Kourkoutis, Y. Hikita, and H.Y. Hwang, "Synthesis of freestanding single-crystal perovskite films and heterostructures by etching of sacrificial water-soluble layers," (2016).

[23] F.M. Chiabrera, S. Yun, Y. Li, R.T. Dahm, H. Zhang, C.K.R. Kirchert, D. V Christensen, F. Trier, T.S. Jespersen, and N. Pryds, "Freestanding Perovskite Oxide Films: Synthesis, Challenges, and Properties," Ann Phys **534**(9), 2200084 (2022).

[24] D. Lu, D.J. Baek, S.S. Hong, L.F. Kourkoutis, Y. Hikita, and H.Y. Hwang, "Synthesis of freestanding single-crystal perovskite films and heterostructures by etching of sacrificial water-soluble layers," Nat Mater **15**(12), 1255–1260 (2016).

[25] R. Qiu, B. Peng, H. Liu, Y. Guo, H. Tang, Z. Zhou, and M. Liu, "Epitaxial growth of pure Sr3Al2O6 sacrificial layer for high quality freestanding single-crystalline oxide membranes," Thin Solid Films **773**, 139820 (2023).

[26] Y. Guo, B. Peng, R. Qiu, G. Dong, Y. Yao, Y. Zhao, Z. Zhou, and M. Liu, "Self-Rolling-Up Enabled Ultrahigh-Density Information Storage in Freestanding Single-Crystalline Ferroic Oxide Films," Adv Funct Mater **33**(20), 2213668 (2023).

[27] C. Xu, X. Wu, G. Huang, and Y. Mei, "Rolled-up Nanotechnology: Materials Issue and Geometry Capability," Adv Mater Technol **4**(1), 1800486 (2019).

[28] L.D. Landau, E.M. Lifshitz, A.M. Kosevich, and L.P. Pitaevskii, *Theory of Elasticity: Volume 7* (Elsevier, 1986).

[29] M.N. Iliev, M. V Abrashev, H.-G. Lee, V.N. Popov, Y.Y. Sun, C. Thomsen, R.L. Meng, and C.W. Chu, "Raman spectroscopy of orthorhombic perovskitelike $\mathrm{YMnO}_{3}$ and $\mathrm{LaMnO}_{3}$," Phys Rev B **57**(5), 2872–2877 (1998).

[30] P.J. Graham, P. Rovillain, M. Bartkowiak, E. Pomjakushina, K. Conder, M. Kenzelmann, and C. Ulrich, "Spin-phonon and magnetoelectric coupling in oxygen-isotope substituted $\mathrm{TbMnO}_{3}$ investigated by Raman scattering," Phys Rev B **105**(17), 174438 (2022).

[31] L. Martín-Carrón, J. Sánchez-Benítez, and A. de Andrés, "High-pressure dependence of Raman phonons of RMnO3 (R=Pr,Tb)," J Solid State Chem **171**(1), 313–316 (2003).

[32] A. Kumar, P. Shahi, S. Kumar, K.K. Shukla, R.K. Singh, A.K. Ghosh, A.K. Nigam, and S. Chatterjee, "Raman effect and magnetic properties of doped TbMnO3," J Phys D Appl Phys **46**(12), 125001 (2013).

[33] D.A. Mota, A. Almeida, V.H. Rodrigues, M.M.R. Costa, P. Tavares, P. Bouvier, M. Guennou, J. Kreisel, and J.A. Moreira, "Dynamic and structural properties of orthorhombic rare-earth manganites under high pressure," Phys Rev B **90**(5), 54104 (2014).

[34] P. Rovillain, M. Cazayous, Y. Gallais, A. Sacuto, M.-A. Measson, and H. Sakata, "Magnetoelectric excitations in multiferroic $\text{TbMnO}_{3}$ by Raman scattering," Phys Rev B **81**(5), 54428 (2010).